\documentclass[fleqn,usenatbib]{mnras}

\usepackage{newtxtext,newtxmath}

\usepackage[T1]{fontenc}
\usepackage{ae,aecompl}

\usepackage{graphicx}	
\usepackage{amsmath}	
\usepackage{amssymb}	
\usepackage{units}
\usepackage{float}

\title[Massive Stars \&\ GMCs]{The Elephant in the Room: The Importance of Where and When Massive Stars Form in Molecular Clouds}

\author[Grudi\'{c} \& Hopkins]{
Michael Y. Grudi\'{c}$^{1}$, Philip F. Hopkins$^{1}$
\\
$^{1}$TAPIR, Mailcode 350-17, California Institute of Technology, Pasadena, CA 91125, USA\\
}

\date{Accepted XXX. Received YYY; in original form ZZZ}

\pubyear{2018}
\hypersetup{draft}
\begin{document}
\label{firstpage}
\pagerange{\pageref{firstpage}--\pageref{lastpage}}
\maketitle
\begin{abstract}
Most simulations of galaxies and massive giant molecular clouds (GMCs) cannot explicitly resolve the formation (or predict the main-sequence masses) of {\em individual} stars. So they must use some prescription for the amount of feedback from an {\em assumed} population of massive stars (e.g.\ sampling the initial mass function [IMF]). We perform a methods study of simulations of a star-forming GMC with stellar feedback from UV radiation, varying only the prescription for determining the luminosity of each stellar mass element formed (according to different IMF sampling schemes). We show that different prescriptions can lead to widely varying (factor of $\sim 3$) star formation efficiencies (on GMC scales) even though the average mass-to-light ratios agree. Discreteness of sources is important: radiative feedback from fewer, more-luminous sources has a greater effect for a given total luminosity. These differences can dominate over other, more widely-recognized differences between similar literature GMC-scale studies (e.g. numerical methods, cloud initial conditions, presence of magnetic fields). Moreover the differences in these methods are not purely numerical: some make different implicit assumptions about where and how massive stars form, and this remains deeply uncertain in star formation theory.
\end{abstract}




\section{Introduction}
Massive stars are rare, but their radiation, winds, and supernova explosions dominate the energy liberated from a stellar population. It is thought that feedback from massive stars is a crucial element for regulating star formation on scales ranging from entire galaxies to individual star clusters \citep{mckee:2007.review, naab.ostriker:2017.review}. In the latter case, significant theoretical efforts have been devoted to understanding how feedback from massive stars sets the star formation efficiency (SFE) of star-forming giant molecular clouds (GMCs), the fraction of the initial gas mass that is converted to stars before feedback disrupts the cloud and halts star formation. An understanding of the SFE of GMCs is important for understanding the origins of the star cluster mass function and its connection to the GMC mass function \citep{elmegreen:1997.open.closed.cluster.same.mf.form,fall:2010.sf.eff.vs.surfacedensity}, the origins of gravitationally bound globular clusters \citep{hills:1980,baumgardt.kroupa:2007,kruijssen:2012.cluster.formation.efficiency}, and the distribution and pre-conditioning of supernova explosions, which affects the efficiency of stellar feedback on galactic scales \citep{keller:2014.superbubbles,fielding:2018.superbubbles}.

Significant progress has been made on this problem as the necessary computational techniques have become available. Many numerical experiments have been performed in which a self-gravitating molecular cloud is evolved in isolation, subject to self-gravity, hydrodynamics, stellar feedback, and possibly detailed cooling and chemistry physics (\citealt{murray:molcloud.disrupt.by.rad.pressure, vazquez:2010, dale:2012, dale:2013.momwinds, colin:2013, dale:2014, skinner:2015.ir.molcloud.disrupt, raskutti:2016.gmcs,howard:2016, howard:2017,vazquez:2017, dale:2017, kim:2017.rhd, gavagnin:2017.rhd.cluster.formation,grudic:2016.sfe,kim:2018}, for review see \citealt{krumholz:2014.feedback.review,dale:2015.review}). For GMC properties consistent with those found in the local Universe ($\Sigma_{gas} \sim \unit[50]{M_\odot \,pc^{-2}}$, $M \sim \unit[10^4-10^6]{M_\odot}$, \citealt{bolatto:2008.gmc.properties}), the most important feedback mechanism for regulating star formation on GMC scales is generally agreed to be UV photons from massive stars. UV photons heat and ionize the ISM upon absorption by gas or dust, while also imparting momentum upon absorption, creating expanding HII regions that ultimately unbind the remaining gas in the cloud. However, theoretical consensus on the specific SFE at which cloud disruption occurs (or even whether it occurs at all, \citealt{howard:2016}) has been slower to develop. As an extreme example, \citet{raskutti:2016.gmcs} simulated a molecular cloud of initial mass $5\times 10^4 M_\odot$ and radius $15 {\rm pc}$ and obtained a SFE of $\sim 40\%$, while \citet{grudic:2016.sfe} simulated the same cloud model with nominally the same feedback physics and obtained $\sim 4\%$, an order of magnitude smaller.

Discrepancies are not necessarily surprising when one considers the compounded variations that can arise when using different hydrodynamics methods, sink particle prescriptions, and perhaps most importantly, radiative transfer approximations. Variations due to these numerical details warrant some exploration, as studies that compare radiative transfer methods while controlling for other factors are few, and none are exhaustive. \citet{raskutti:2016.gmcs} performed simulations treating the effects of photon momentum (ie. radiation pressure) from UV photon absorption with an M1 radiation transfer scheme \citep{skinner:2013.hyperion}, which \citet{kim:2017.rhd} subsequently compared to adaptive ray tracing results using otherwise the same code, and found a SFE a factor of $\sim 2$ smaller. \citet{hopkins.grudic:2018.rp} also performed GMC simulations comparing the ray-based {\small LEBRON} radiative transfer approximation \citep{fire2} with an M1 scheme \citep{rosdahl:2015.rhd}, and also found agreement at the factor of $\sim 2$ level. Therefore, variations in radiative transfer techniques can likely account for some of the variation found in the literature, but probably not all of it. This motivates the consideration of other factors.

Several of the studies cited above compared additional physics (e.g.\ including or ignoring magnetic fields), or varying the cloud initial conditions (e.g.\ considering clouds with or without pre-initialized fully-developed turbulence, with or without significant rotational support, and with or without a global density profile): the general conclusion is that these, too, can influence the predicted star formation efficiency by at most a factor $\sim 2$ (see references above and \citealt{klessen:2000.cloud.collapse,krumholz:2011.rhd.starcluster.sim,price.bate:2008.magnetic.fields}). Others have shown that including or excluding other sources of feedback besides UV radiation alone, e.g.\ O/B stellar winds (which carry a similar momentum flux to the UV radiation field), can have a similar effect.

In \citet{hopkins.grudic:2018.rp} we argued that another potential error source can arise when using the most common method for coupling radiation pressure to gas, which underestimates the imparted momentum from a point source if the photon mean free path is smaller than the fluid resolution. \citet{krumholz:2018.rp} subsequently pointed out another previously-overlooked numerical pitfall: photon absorption around an accreting massive star can occur deep in the potential well on scales smaller than the resolution limit of most simulations, preventing it from imparting momentum on larger scales. They argued that the failure to resolve this effect could also explain some of the discrepancy, and proposed a subgrid model to approximate this effect in numerical simulations. 

This led us to consider the broader important question that we will address here: how do the details of how the {\it sources} of stellar feedback are modeled in simulations affect the cloud-scale SFE? Clearly, when simulations attempt to model the formation of massive stars self-consistently, the details of the IMF will become important for feedback, as UV luminosity is a steep function of stellar mass. However, most GMC-scale and all galaxy-scale hydrodynamics simulations either lack the resolution or the physical realism to do this self-consistently, so feedback is often treated with phenomenological prescription, {\it assuming} an underlying stellar mass distribution that is being sampled in some manner. In this work we will compare several such techniques, and determine the effect of these numerical choices upon the cloud-scale SFE in simulations.




\section{Simulations}

\subsection{Numerical methods}

We simulate an isolated turbulent molecular cloud with {\small GIZMO}, a multi-physics N-body and hydrodynamics code \citep{hopkins:gizmo}\footnote{\url{http://www.tapir.caltech.edu/~phopkins/Site/GIZMO.html}}. We solve the equations of hydrodynamics with the Lagrangian Meshless Finite Mass (MFM) method. We account for a wide range of ISM heating and cooling physics, using the rates and implementations used in the FIRE-2 simulations \citep{fire2}\footnote{\url{https://fire.northwestern.edu/}}. Star formation is treated with an accreting sink particle method described in \citet{guszejnov:2018.isothermal}, which uses multiple checks for sink formation and accretion, similar to \citet{Federrath_2010_sink_particle}. For simplicity, we consider only the effects of feedback due to the absorption of UV photons from stars, accounting for the effects of photo-heating and radiation pressure as in \citet{fire2}.

\subsection{Initial conditions}
We replicate the initial conditions of the fiducial cloud model in \citet{kim:2018}, a GMC with a top-hat density profile with mass $10^5M_\odot$ and radius $\unit[20]{pc}$. The initial velocity field is a solenoidal Gaussian random field with power spectrum $|\mathbf{\tilde{v}}\left(\mathbf{k}\right)|^2 \propto k^{-4}$ \citep{gammie.ostriker:1996.mhd.dissipation}, normalized so that the initial kinetic energy is equal to the gravitational potential energy. The initial metallicity of the cloud is normalized to solar abundances, accounted for self-consistently in the cooling function and the dust opacity to UV photons as in \cite{fire2}. In all simulations we resolve the initial gas mass in $128^3$ Lagrangian gas cells, for a mass resolution of $\unit[0.048]{M_\odot}$. Initial conditions were generated with the {\small MakeCloud} code\footnote{\url{https://github.com/omgspace/MakeCloud}}.

\subsection{IMF sampling models}
\label{section:prescriptions}
We perform simulations with a range of different prescriptions for the specific bolometric and ionizing luminosities assigned to the stellar mass elements (ie. sink particles) in the simulation. These are all intended to mimic certain aspects of the effect of sampling a finite number of stars from an underlying probability distribution function (ie. the IMF). Each recovers the same net specific luminosities in the limit $M_\star >> 10^{3}$, but each approaches that limit in a different manner as stars form in the simulation. All of these prescriptions have advantages and disadvantages -- in this work we remain agnostic about the relative physical realism of these models, which is difficult to evaluate without a self-consistent treatment of the physics of massive star formation. We consider only models that work under the assumption that the IMF can indeed by interpreted as probability distribution to be sampled from until a given stellar mass reservoir is exhausted. This must break down at some level, due to mass conservation if no other reason. However, the details of how the IMF emerges are poorly understood, and the stochastic sampling hypothesis is consistent with current observations \citep{bastian:2010.imf, fumagalli:2011.imf,offner:2014.imf.review}.

\subsubsection{\texttt{IMFMEAN}: Simple IMF-averaging}
The simplest approach is to assume that all stellar mass elements 
in the simulation have the same specific luminosity as a well-sampled IMF, which for a very young stellar population with age $\ll 3.5\,$Myr and a \citet{kroupa:imf} IMF is approximately
\begin{equation}
\left \langle \frac{L_\star}{M_\star}\right \rangle_{IMF} = 1140 L_\odot M_\odot^{-1}
\end{equation}
This is approximately constant until $t \approx 3.5\,$Myr, then decreases appropriately as massive stars die according to an adopted stellar evolution model \citep[e.g.][]{starburst99}. 
The well-sampled assumption is expected to be valid in systems where the total stellar mass is $>> 10^3 M_\odot$, and is a common choice for galaxy simulations that might not even resolve mass scales smaller than this \citep{hopkins:2011.mom.feedback,agertz:2013.new.stellar.fb.model,fire2}, although it has also been used in smaller-scale cluster formation simulations \citep{grudic:2016.sfe,grudic:2017,hopkins.grudic:2018.rp, kim:2018.fire.gcs}. This is the only prescription that guarantees that the specific luminosity is always equal to the ensemble over all possible IMF samplings. However, this is not necessarily desirable in all problems. The method has a serious drawback in the regime of low-mass star cluster formation: when sampling an IMF from a small reservoir of stellar mass, most realizations sample no massive stars at all. Therefore, the mean specific luminosity is due to those very few possible samplings that do contain massive stars and have specific luminosities much greater than the mean. The effect of this is to give a specific luminosity that is much larger than the vast majority of possible realizations of low-mass clusters, and much less than those few realizations that do, averaging out a major source of physically-real stochasticity.

In addition to the standard \texttt{IMFMEAN} scheme, we consider a variant supplemented by the subgrid model introduced by \citet{krumholz:2018.rp}, \texttt{IMFMEAN-K18}. To mimic the effect of photon absorption in a dust destruction front at unresolved scales, we simply switch off UV feedback from a sink particle when its accretion rate exceeds the threshold value
\begin{equation}
\frac{\dot{M}}{M_\odot\,\rm{yr}^{-1}} > 6.5\times10^{-4} \left(\frac{L}{10^6 L_\odot}\right)^{3/4}.
\label{eq:krumholz2018}
\end{equation}
Because our sink particles accrete discrete Lagrangian gas cells, we apply exponential smoothing to the accretion rate for this check, with an $e$-folding time $\tau_{accr}=10^5 \mathrm{yr}$, motivated by the fiducial timescale for massive star formation. We have experimented with setting this parameter to $10^4\mathrm{yr}$ and $10^6\mathrm{yr}$ and found that it has no important effect on the SFE.

\subsubsection{\texttt{IMFMED}: scaling to a median value}
An alternative approach to using the IMF-averaged mean is to use the median or most likely (which are close) value over the ensemble of IMF samplings, assuming that the total stellar mass formed in the simulation can be interpreted as a coeval stellar population. \citet{kim:2016.dusty.hii} developed this approach by sampling stellar populations with the {\small SLUG} code \citep{slug} for a range of cluster masses and deriving a fitting formula to the median value sampled at each mass scale. The median value is very small for star clusters less than a couple $100M_\odot$, and scales steeply toward the well-sampled value once $M_\star \sim 1000 M_\odot$. This model was used in their subsequent RHD simulations \citep{kim:2017.rhd,kim:2018}, and is the one we implement here.

The \texttt{IMFMED} model will give a value more representative of a ``typical'' sampling. The disadvantage of this approach is that it lacks locality: star formation in one region of the cloud influences the amount of feedback everywhere else, which is unphysical and cannot generalize to more complicated systems in which the very definition of a progenitor cloud, and hence which stars belong to which coeval population, is ill-defined. It and \texttt{IMFMEAN} share another disadvantage: every sink particle has the same light-to-mass ratio, which is artificially smoother than the true distribution of luminous sources in a star cluster. This motivates our next prescription.

\subsubsection{\texttt{POISSON}: Poisson-sampling quantized sources}
To model the discreteness of luminous sources, we can sample a quantized number of `O-stars' in each sink particle, such that the expectation value is still the IMF-averaged value. We adopt the presciption of \citet{su:2017.discreteness}, which assigns to each sink particle a number of `O-stars' sampled from a Poisson distribution, with expectation value
\begin{equation}
\mu = \frac{m_{particle}}{\Delta m},
\end{equation}
where $m_{particle}$ is the mass of the sink particle and $\Delta m$ was taken to be $100M_\odot$ in \citet{su:2017.discreteness}. Then, each `O-star' is taken to have luminosity
\begin{equation}
L = \Delta m \left\langle \frac{L_\star}{M_\star}\right \rangle_{IMF}.
\end{equation}
This technique has the advantage of giving a more realistic {\it number} of feedback-injecting sources for a given amount of stellar mass. It also captures the effect of under-sampling the IMF, but stochastically rather than causally as \texttt{IMFmed}. Although the version used in \citet{su:2017.discreteness} sampled only one species of `O-star', it is in principle extensible to an arbitrary number of species \citep{sormani:2016.imf.sampling}. The details of how the luminosity is discretized, ie. few sources versus many, is potentially important. Feedback from a single luminous source might be more efficient than that of many smaller sources because it is more concentrated and less subject to momentum cancellation \citep[e.g.][]{dale:2017,kim:2018}. On the other hand, it could also be effectively weaker because luminous sources are only likely to appear once a certain amount of stellar mass has formed, at which point collapse may be more advanced and the resulting structure more difficult to disrupt. We consider two variants of this prescription, with $\Delta m=100M_\odot$ and 
$\Delta m=1000M_\odot$, denoted \texttt{POISSON100} and \texttt{POISSON1000} respectively.
\\
\\

A notable omission from this section is the prescription of \citet{howard:2016}, which interprets each sink particle as an individual cluster, and effectively applies a variant of the \texttt{IMFMED} prescription to each of these clusters individually. We have experimented with this prescription and found it to be numerically problematic because the characteristic mass of sink particles drops as a function of mass resolution, as has generally been found in other simulations \citep{bate:2009.imf, federrath:2017.imf,guszejnov:2018.isothermal}. Thus in the limit of high resolution, feedback is made effectively weaker, and numerical convergence in the SFE is not achieved. This type of prescription can only converge for sink particle algorithms that imprint a characteristic size or density scale other than the numerical resolution, which requires certain assumptions about the nature of star cluster formation that we will not consider here.

\section{Results}
\begin{figure*}
\includegraphics[width=\textwidth]{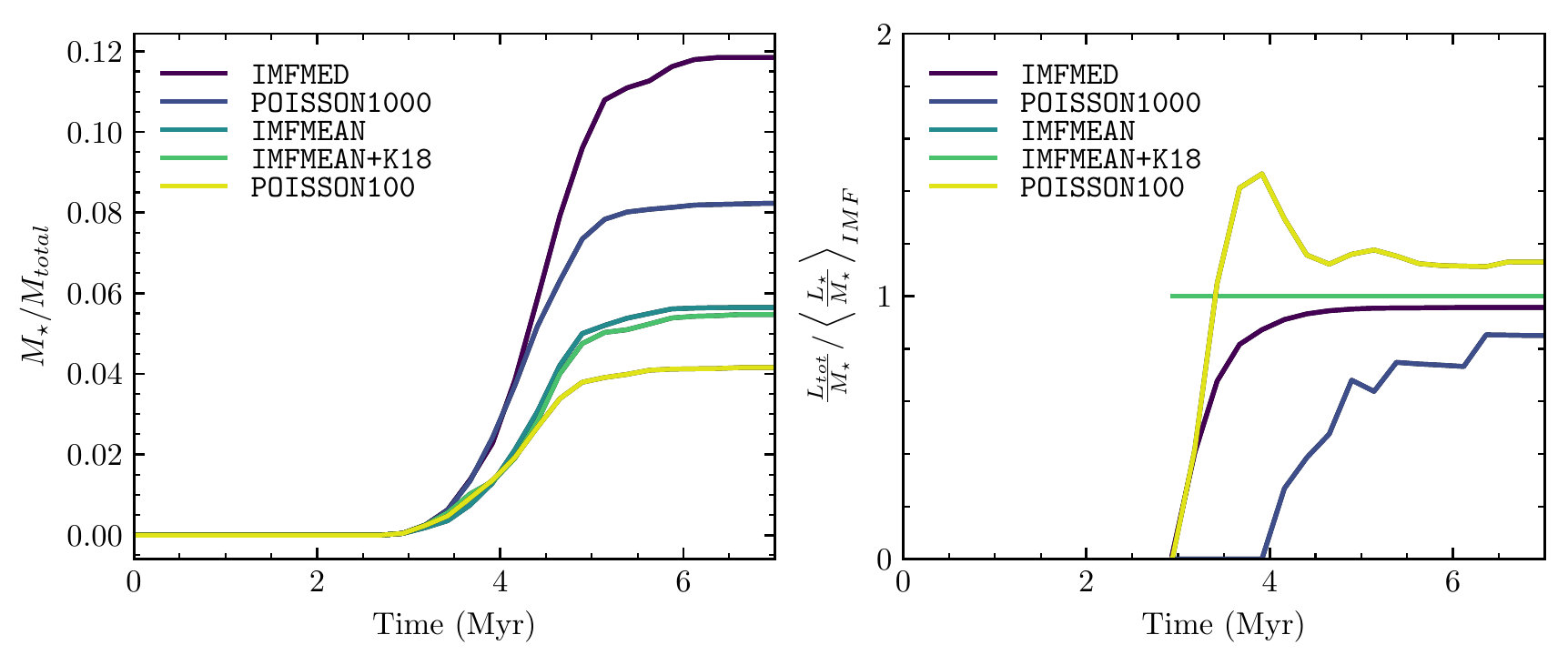}
\caption{{\it Left}: Integrated star formation efficiency (fraction of cloud mass converted to stars) in the simulations as a function of time, for simulations run with each of the different subgrid feedback models considered (\S\ref{section:prescriptions}). {\it Right}: bulk stellar light-to-mass ratio according to the different prescriptions, normalized to the IMF-averaged value. Note that the \texttt{IMFMEAN-K18} curve does not include the effect of turning off feedback according to Equation \ref{eq:krumholz2018}, but we find that the effect is small (see discussion in \S\ref{section:results}). The different prescriptions all approach the same IMF-averaged value in the limit $M_\star >> 10^3 M_\odot$, but they differ in how they approach this limit.}
\label{fig}
\end{figure*}
\label{section:results}

The simulated clouds evolve according to the usual sequence of events found in this type of simulation \citep[e.g.][]{grudic:2016.sfe}: turbulence dissipates in shocks on the crossing timescale, and the cloud collapses into dense substructures that eventually form stars. Eventually, the cumulative effect of stellar feedback is sufficient to disrupt the cloud, halting star formation. In Figure \ref{fig} we plot the integrated star formation efficiency $M_\star /M$ and the light-to-mass ratio as a function of time for each of the five different prescriptions used.

The SFE varies considerably with the feedback prescription used: \texttt{IMFMED} ended with a SFE of $12\%$, while $\texttt{POISSON100}$ gave $4\%$, with the others lying in between. This is despite the fact that the final light-to-mass ratios from each prescription all agreed within 10\%, as at least $4000M_\odot$ forms in each simulation. We therefore find that the details of IMF sampling prescriptions for feedback can have a considerable effect on the SFE of simulated molecular clouds. In particular, we find that the \texttt{IMFMEAN} value of $5\%$ is reasonably consistent with \citet{grudic:2016.sfe}, which used that prescription, while \texttt{IMFMED} gives a SFE of $12\%$, consistent with \citet{kim:2018}, explaining the discrepancy between those specific works.

The \texttt{IMFMEAN+K18} prescription gives results that are nearly indistinguishable from the standard \texttt{IMFMEAN} prescription, despite the fact that it always gives less feedback. We have generally found that the fraction of time during which the criteria for turning off feedback (Equation \ref{eq:krumholz2018}) are satisfied is very short compared to the lifetime of the GMC. Star particles accrete rapidly out of dense cores, and accretion halts either when the gas is exhausted or when the star particle is dynamically ejected out of its natal clump and into a void. Once Equation \ref{eq:krumholz2018} is no longer satisfied, feedback turns on and generally drives an outflow around the star. Once this outflow has been initiated, it tends not to end. Therefore even a brief lapse in the accretion rate can effectively end the accretion history.

Even assuming an infinite reservoir for accretion, an upper bound on the amount of time that Equation \ref{eq:krumholz2018} can apply can be derived from the observed properties massive stars. To maximize this time, we assume that Equation \ref{eq:krumholz2018} holds as an equality. The luminosity of stars more massive than $\sim 20 M_\odot$ is:
\begin{equation}
\frac{L}{10^6 L_\odot} \approx 0.03 \frac{M}{M_\odot}.
\end{equation}
Substituting this into Equation \ref{eq:krumholz2018} gives
\begin{equation}
\frac{\dot{M}}{M_\odot\,\rm{yr}^{-1}} = 4.7 \times 10^{-5} \left(\frac{M}{M_\odot}\right)^{3/4}.
\end{equation}

Over the mass range of massive stars, the solution to this equation is well-approximated by exponential growth with an $e$-folding time of $\unit[40]{kyr}$, so within a few $\unit[100]{kyr}$ the maximum stellar mass on the order of $\unit[100]{M_\odot}$ must be reached. Because this is much shorter than even the shortest GMC lifetimes, the effect upon the cloud-scale SFE is small. However, we emphasize that the prescription could easily have more important effects on smaller scales or shorter timescales, such as influencing the accretion history of individual protostars or the formation of a dense star cluster.

\section{Discussion}
We have shown that when simulating the evolution of an isolated molecular cloud, the specific prescription for massive stellar feedback used can affect the star formation efficiency of the cloud (and by extension, the properties of the star cluster formed) at least at the factor of $\sim 3$ level. This is despite the fact that all simulations eventually form at least several $10^3 M_\odot$ in stars, so the IMFs in all cases are well-sampled and the final light-to-mass ratios do not differ widely.

The simplest analytic estimate of the feedback-regulated SFE of a molecular cloud can be obtained by simply equating the bulk momentum injection rate due to feedback to the weight of the cloud due to self-gravity. In the limit of small SFE \citep{fall:2010.sf.eff.vs.surfacedensity,murray:molcloud.disrupt.by.rad.pressure,hopkins:fb.ism.prop,grudic:2016.sfe,kim:2018}:
\begin{equation}
\rm{SFE} \propto \Sigma_{gas}/\frac{L_\star}{M_\star},
\label{eq:forcebalance}
\end{equation}
where $\Sigma_{gas}$ is the mean surface density of the cloud. If this force balance is assumed to hold at the time of cloud disruption, then we would expect that the variation in SFE would not exceed the variation in $\frac{L_\star}{M_\star}$, but the simulations show that this is not the case: all simulations end with the same $\frac{L_\star}{M_\star}$ within $10\%$, yet the variation in SFE is a factor of 3.

We generally find that prescriptions that take longer to approach the fully-sampled specific luminosity have SFE that can be a factor 2-3 higher than the fiducial \texttt{IMFMEAN} prescription. The physical reason for this is of course that the efficiency of feedback does not depend only upon the bulk ionization or momentum deposition rate: it also depends on {\it where} and {\it when} the absorption event occurs, a point deftly illustrated in recent work \citep{jumper.matzner:2018.rp,krumholz:2018.rp}. Specifically, recombination and cooling times are shorter at higher density, suppressing radiative heating effects, while momentum imparted in a deeper potential well provides less terminal momentum, and if the well is sufficiently deep the momentum might not be sufficient to launch a wind at all.

This raises a point that is more broadly important: the effectiveness of feedback from massive stars depends on much more than just the bulk light-to-mass ratio arising from the IMF -- it depends on when and under what conditions massive stars form. This should hold quite generally, so although we have only considered schemes for injecting feedback from an assumed IMF, this has implications for calculations that attempt to resolve the IMF self-consistently. The particulars of where massive stars form in the cloud, when they form relative to other stars, and how long they take to form should all influence the behaviour of stellar feedback. The resulting influence on feedback influences the evolution of the entire cloud and the stellar population that will form.

Counter-intuitively, the \texttt{POISSON1000} simulation had lower SFE than the \texttt{IMFMED} simulation despite the fact that its light-to-mass ratio was lower at all times. This implies that feedback in the \texttt{POISSON1000} was more effective for a given specific luminosity. The effect is due to the different discretizations of luminosity among the sink particles: with \texttt{IMFMED}, all sink particles have the same specific luminosity, while for \texttt{POISSON1000} the luminosity was concentrated in only five sources at the time star formation ended. Therefore, radiative feedback from fewer, more-luminous sources is more efficient, a result analogous to what has been found for the clustering of supernova explosions \citep{keller:2014.superbubbles,fielding:2018.superbubbles}. We can conjecture that the true IMF-resolved solution is probably closer to the discrete limit, because the bolometric and especially the ionizing luminosity will generally be dominated by the few most-massive stars.

We note that similar experiments to those shown here were considered on a {\em galactic} scale in \citep{su:2017.discreteness}, who argued that galaxy-averaged quantities (e.g.\ stellar masses, sizes, morphologies, abundance patterns, statistics of their star formation histories) were not strongly influenced by the IMF sampling scheme. This is not surprising, as the spatial and time scales of self-regulation via feedback in those simulations are much longer ($\gg 10\,$Myr), so most of the dynamics occurs in the well-sampled IMF limit (even in dwarf galaxies). Moreover other studies have shown that even artificially raising or lowering the GMC-scale star formation efficiency by much larger factors than those seen here produces relatively weak effects on galactic properties, because of global self-regulation by outflows and pressure balance in the ISM \citep{hopkins:2011.mom.feedback,agertz:2015.sf.fb,cafg:sf.fb.reg.kslaw,orr:2018.kennicutt.schmidt}. However, our study here suggests that sub-galactic but still large-scale quantities, e.g.\ properties of star clusters and lifetime/mass of molecular gas at any given time, could be significantly influenced by the physics discussed here.

It is of course possible to develop more sophisticated IMF-sampling schemes \citep[for examples, see][]{hu:2016, fujimoto:2018, emerick:2018.methods}, coupled to more detailed stellar evolution models for feedback, and this can provide some improvements for coarse-grained IMF prescriptions (especially for phenomena like SNe occurring on much longer timescales). However, we stress that on the spatial and time scales of GMCs, this is not obviously ``more correct'': the real issue is not the statistical method by which the IMF is sampled. Rather, it is the fact that these (and all of the methods discussed here) are fundamentally assigning the question of where and when massive stars form to a ``sub-grid'' model, which does not know about the local (resolved) conditions in the GMC/ISM. Most of the stellar mass will form wherever nature can form a $\sim 0.1\,M_{\odot}$ star -- but low-mass cores in low-density environments almost certainly cannot form the $\gtrsim 40\,M_{\odot}$ stars that dominate the UV production in a massive cluster. And allowing massive stars to form ``stochastically'' in such environments may likely over-estimate their effects. It is also not obvious that neglecting the accretion/formation and protostellar/pre-main sequence evolution of such stars is a valid approximation on the $\sim 1\,$Myr timescales of interest here.

As such, what we have shown is that significant, intrinsic uncertainties clearly still exist about the effects of stellar feedback at the GMC scale, at least at the level demonstrated here. These uncertainties will remain until the emergence of the IMF from GMC dynamics is understood in a self-consistent framework. Sub-grid feedback prescriptions should ultimately be informed by simulations that are able to follow the formation of a stellar population at the level of resolution required to model the formation of individual massive stars in an accurate and robust manner, so that one can model in a physically-motivated manner when and where in a simulation massive stars form.

\section*{Acknowledgements}
We thank Mark Krumholz, Stella Offner, Claude-Andr\'{e} Faucher-Gigu\`{e}re, Norman Murray, Eliot Quataert, James Dale, Ian Bonnell, Hui Li, Jeong-Gyu Kim, Eve Ostriker, Michael S. Fall, Christopher Matzner, Benny Tsang, and Milos Milosavljevi\'{c} for enlightening discussions that informed and motivated this work. Support for MYG and PFH was provided by a James A Cullen Memorial Fellowship, an Alfred P. Sloan Research Fellowship, NSF Collaborative Research Grant \#1715847 and CAREER grant \#1455342, and NASA grants NNX15AT06G, JPL 1589742, 17-ATP17-0214. Numerical calculations were run on the Caltech compute cluster ``Wheeler,'' allocations from XSEDE TG-AST130039 and PRAC NSF.1713353 supported by the NSF, and NASA HEC SMD-16-7592. This research has made use of the NASA Astrophysics Data System, {\texttt ipython} \citep{ipython}, {\texttt numpy}, {\texttt scipy} \citep{scipy}, {\texttt numba} \citep{numba}, and {\texttt matplotlib} \citep{matplotlib}. 




\bibliographystyle{mnras}
\bibliography{master} 

\begin{thebibliography}{}
\makeatletter
\relax
\def\mn@urlcharsother{\let\do\@makeother \do\$\do\&\do\#\do\^\do\_\do\%\do\~}
\def\mn@doi{\begingroup\mn@urlcharsother \@ifnextchar [ {\mn@doi@}
  {\mn@doi@[]}}
\def\mn@doi@[#1]#2{\def\@tempa{#1}\ifx\@tempa\@empty \href
  {http://dx.doi.org/#2} {doi:#2}\else \href {http://dx.doi.org/#2} {#1}\fi
  \endgroup}
\def\mn@eprint#1#2{\mn@eprint@#1:#2::\@nil}
\def\mn@eprint@arXiv#1{\href {http://arxiv.org/abs/#1} {{\tt arXiv:#1}}}
\def\mn@eprint@dblp#1{\href {http://dblp.uni-trier.de/rec/bibtex/#1.xml}
  {dblp:#1}}
\def\mn@eprint@#1:#2:#3:#4\@nil{\def\@tempa {#1}\def\@tempb {#2}\def\@tempc
  {#3}\ifx \@tempc \@empty \let \@tempc \@tempb \let \@tempb \@tempa \fi \ifx
  \@tempb \@empty \def\@tempb {arXiv}\fi \@ifundefined
  {mn@eprint@\@tempb}{\@tempb:\@tempc}{\expandafter \expandafter \csname
  mn@eprint@\@tempb\endcsname \expandafter{\@tempc}}}

\bibitem[\protect\citeauthoryear{{Agertz} \& {Kravtsov}}{{Agertz} \&
  {Kravtsov}}{2015}]{agertz:2015.sf.fb}
{Agertz} O.,  {Kravtsov} A.~V.,  2015, \mn@doi [\apj]
  {10.1088/0004-637X/804/1/18}, \href
  {http://adsabs.harvard.edu/abs/2015ApJ...804...18A} {804, 18}

\bibitem[\protect\citeauthoryear{{Agertz}, {Kravtsov}, {Leitner}  \&
  {Gnedin}}{{Agertz} et~al.}{2013}]{agertz:2013.new.stellar.fb.model}
{Agertz} O.,  {Kravtsov} A.~V.,  {Leitner} S.~N.,   {Gnedin} N.~Y.,  2013,
  \mn@doi [\apj] {10.1088/0004-637X/770/1/25}, \href
  {http://adsabs.harvard.edu/abs/2013ApJ...770...25A} {770, 25}

\bibitem[\protect\citeauthoryear{{Bastian}, {Covey}  \& {Meyer}}{{Bastian}
  et~al.}{2010}]{bastian:2010.imf}
{Bastian} N.,  {Covey} K.~R.,   {Meyer} M.~R.,  2010, \mn@doi [\araa]
  {10.1146/annurev-astro-082708-101642}, \href
  {http://adsabs.harvard.edu/abs/2010ARA%26A..48..339B} {48, 339}

\bibitem[\protect\citeauthoryear{{Bate}}{{Bate}}{2009}]{bate:2009.imf}
{Bate} M.~R.,  2009, \mn@doi [\mnras] {10.1111/j.1365-2966.2008.14165.x}, \href
  {http://adsabs.harvard.edu/abs/2009MNRAS.392.1363B} {392, 1363}

\bibitem[\protect\citeauthoryear{{Baumgardt} \& {Kroupa}}{{Baumgardt} \&
  {Kroupa}}{2007}]{baumgardt.kroupa:2007}
{Baumgardt} H.,  {Kroupa} P.,  2007, \mn@doi [\mnras]
  {10.1111/j.1365-2966.2007.12209.x}, \href
  {http://adsabs.harvard.edu/abs/2007MNRAS.380.1589B} {380, 1589}

\bibitem[\protect\citeauthoryear{{Bolatto}, {Leroy}, {Rosolowsky}, {Walter}  \&
  {Blitz}}{{Bolatto} et~al.}{2008}]{bolatto:2008.gmc.properties}
{Bolatto} A.~D.,  {Leroy} A.~K.,  {Rosolowsky} E.,  {Walter} F.,   {Blitz} L.,
  2008, \mn@doi [\apj] {10.1086/591513}, \href
  {http://adsabs.harvard.edu/abs/2008ApJ...686..948B} {686, 948}

\bibitem[\protect\citeauthoryear{{Col{\'{\i}}n}, {V{\'a}zquez-Semadeni}  \&
  {G{\'o}mez}}{{Col{\'{\i}}n} et~al.}{2013}]{colin:2013}
{Col{\'{\i}}n} P.,  {V{\'a}zquez-Semadeni} E.,   {G{\'o}mez} G.~C.,  2013,
  \mn@doi [\mnras] {10.1093/mnras/stt1409}, \href
  {http://adsabs.harvard.edu/abs/2013MNRAS.435.1701C} {435, 1701}

\bibitem[\protect\citeauthoryear{{Dale}}{{Dale}}{2015}]{dale:2015.review}
{Dale} J.~E.,  2015, \mn@doi [\nar] {10.1016/j.newar.2015.06.001}, \href
  {http://adsabs.harvard.edu/abs/2015NewAR..68....1D} {68, 1}

\bibitem[\protect\citeauthoryear{{Dale}}{{Dale}}{2017}]{dale:2017}
{Dale} J.~E.,  2017, \mn@doi [\mnras] {10.1093/mnras/stx028}, \href
  {http://adsabs.harvard.edu/abs/2017MNRAS.467.1067D} {467, 1067}

\bibitem[\protect\citeauthoryear{{Dale}, {Ercolano}  \& {Bonnell}}{{Dale}
  et~al.}{2012}]{dale:2012}
{Dale} J.~E.,  {Ercolano} B.,   {Bonnell} I.~A.,  2012, \mn@doi [\mnras]
  {10.1111/j.1365-2966.2012.21205.x}, \href
  {http://adsabs.harvard.edu/abs/2012MNRAS.424..377D} {424, 377}

\bibitem[\protect\citeauthoryear{{Dale}, {Ngoumou}, {Ercolano}  \&
  {Bonnell}}{{Dale} et~al.}{2013}]{dale:2013.momwinds}
{Dale} J.~E.,  {Ngoumou} J.,  {Ercolano} B.,   {Bonnell} I.~A.,  2013, \mn@doi
  [\mnras] {10.1093/mnras/stt1822}, \href
  {http://adsabs.harvard.edu/abs/2013MNRAS.436.3430D} {436, 3430}

\bibitem[\protect\citeauthoryear{{Dale}, {Ngoumou}, {Ercolano}  \&
  {Bonnell}}{{Dale} et~al.}{2014}]{dale:2014}
{Dale} J.~E.,  {Ngoumou} J.,  {Ercolano} B.,   {Bonnell} I.~A.,  2014, \mn@doi
  [\mnras] {10.1093/mnras/stu816}, \href
  {http://adsabs.harvard.edu/abs/2014MNRAS.442..694D} {442, 694}

\bibitem[\protect\citeauthoryear{{Elmegreen} \& {Efremov}}{{Elmegreen} \&
  {Efremov}}{1997}]{elmegreen:1997.open.closed.cluster.same.mf.form}
{Elmegreen} B.~G.,  {Efremov} Y.~N.,  1997, \mn@doi [\apj] {10.1086/303966},
  \href {http://adsabs.harvard.edu/abs/1997ApJ...480..235E} {480, 235}

\bibitem[\protect\citeauthoryear{{Emerick}, {Bryan}  \& {Mac Low}}{{Emerick}
  et~al.}{2018}]{emerick:2018.methods}
{Emerick} A.,  {Bryan} G.~L.,   {Mac Low} M.-M.,  2018, preprint, \href
  {http://adsabs.harvard.edu/abs/2018arXiv180707182E} {} (\mn@eprint {arXiv}
  {1807.07182})

\bibitem[\protect\citeauthoryear{{Fall}, {Krumholz}  \& {Matzner}}{{Fall}
  et~al.}{2010}]{fall:2010.sf.eff.vs.surfacedensity}
{Fall} S.~M.,  {Krumholz} M.~R.,   {Matzner} C.~D.,  2010, \mn@doi [\apjl]
  {10.1088/2041-8205/710/2/L142}, \href
  {http://adsabs.harvard.edu/abs/2010ApJ...710L.142F} {710, L142}

\bibitem[\protect\citeauthoryear{{Faucher-Gigu{\`e}re}, {Quataert}  \&
  {Hopkins}}{{Faucher-Gigu{\`e}re} et~al.}{2013}]{cafg:sf.fb.reg.kslaw}
{Faucher-Gigu{\`e}re} C.-A.,  {Quataert} E.,   {Hopkins} P.~F.,  2013, \mn@doi
  [\mnras] {10.1093/mnras/stt866}, \href
  {http://adsabs.harvard.edu/abs/2013MNRAS.433.1970F} {433, 1970}

\bibitem[\protect\citeauthoryear{{Federrath}, {Banerjee}, {Clark}  \&
  {Klessen}}{{Federrath} et~al.}{2010}]{Federrath_2010_sink_particle}
{Federrath} C.,  {Banerjee} R.,  {Clark} P.~C.,   {Klessen} R.~S.,  2010,
  \mn@doi [\apj] {10.1088/0004-637X/713/1/269}, \href
  {http://adsabs.harvard.edu/abs/2010ApJ...713..269F} {713, 269}

\bibitem[\protect\citeauthoryear{{Federrath}, {Krumholz}  \&
  {Hopkins}}{{Federrath} et~al.}{2017}]{federrath:2017.imf}
{Federrath} C.,  {Krumholz} M.,   {Hopkins} P.~F.,  2017, in Journal of Physics
  Conference Series. p. 012007, \mn@doi{10.1088/1742-6596/837/1/012007}

\bibitem[\protect\citeauthoryear{{Fielding}, {Quataert}  \&
  {Martizzi}}{{Fielding} et~al.}{2018}]{fielding:2018.superbubbles}
{Fielding} D.,  {Quataert} E.,   {Martizzi} D.,  2018, preprint, \href
  {http://adsabs.harvard.edu/abs/2018arXiv180708758F} {} (\mn@eprint {arXiv}
  {1807.08758})

\bibitem[\protect\citeauthoryear{{Fujimoto}, {Krumholz}  \&
  {Tachibana}}{{Fujimoto} et~al.}{2018}]{fujimoto:2018}
{Fujimoto} Y.,  {Krumholz} M.~R.,   {Tachibana} S.,  2018, \mn@doi [\mnras]
  {10.1093/mnras/sty2132}, \href
  {http://adsabs.harvard.edu/abs/2018MNRAS.480.4025F} {480, 4025}

\bibitem[\protect\citeauthoryear{{Fumagalli}, {da Silva}  \&
  {Krumholz}}{{Fumagalli} et~al.}{2011}]{fumagalli:2011.imf}
{Fumagalli} M.,  {da Silva} R.~L.,   {Krumholz} M.~R.,  2011, \mn@doi [\apj]
  {10.1088/2041-8205/741/2/L26}, \href
  {https://ui.adsabs.harvard.edu/#abs/2011ApJ...741L..26F} {741, L26}

\bibitem[\protect\citeauthoryear{{Gammie} \& {Ostriker}}{{Gammie} \&
  {Ostriker}}{1996}]{gammie.ostriker:1996.mhd.dissipation}
{Gammie} C.~F.,  {Ostriker} E.~C.,  1996, \mn@doi [\apj] {10.1086/177556},
  \href {http://adsabs.harvard.edu/abs/1996ApJ...466..814G} {466, 814}

\bibitem[\protect\citeauthoryear{{Gavagnin}, {Bleuler}, {Rosdahl}  \&
  {Teyssier}}{{Gavagnin} et~al.}{2017}]{gavagnin:2017.rhd.cluster.formation}
{Gavagnin} E.,  {Bleuler} A.,  {Rosdahl} J.,   {Teyssier} R.,  2017, \mn@doi
  [\mnras] {10.1093/mnras/stx2222}, \href
  {http://adsabs.harvard.edu/abs/2017MNRAS.472.4155G} {472, 4155}

\bibitem[\protect\citeauthoryear{{Grudi{\'c}}, {Hopkins},
  {Faucher-Gigu{\`e}re}, {Quataert}, {Murray}  \& {Kere{\v s}}}{{Grudi{\'c}}
  et~al.}{2018a}]{grudic:2016.sfe}
{Grudi{\'c}} M.~Y.,  {Hopkins} P.~F.,  {Faucher-Gigu{\`e}re} C.-A.,  {Quataert}
  E.,  {Murray} N.,   {Kere{\v s}} D.,  2018a, \mn@doi [\mnras]
  {10.1093/mnras/sty035}, \href
  {http://adsabs.harvard.edu/abs/2018MNRAS.475.3511G} {475, 3511}

\bibitem[\protect\citeauthoryear{{Grudi{\'c}}, {Guszejnov}, {Hopkins},
  {Lamberts}, {Boylan-Kolchin}, {Murray}  \& {Schmitz}}{{Grudi{\'c}}
  et~al.}{2018b}]{grudic:2017}
{Grudi{\'c}} M.~Y.,  {Guszejnov} D.,  {Hopkins} P.~F.,  {Lamberts} A.,
  {Boylan-Kolchin} M.,  {Murray} N.,   {Schmitz} D.,  2018b, \mn@doi [\mnras]
  {10.1093/mnras/sty2303}, \href
  {http://adsabs.harvard.edu/abs/2018MNRAS.481..688G} {481, 688}

\bibitem[\protect\citeauthoryear{{Guszejnov}, {Hopkins}, {Grudi{\'c}},
  {Krumholz}  \& {Federrath}}{{Guszejnov}
  et~al.}{2018}]{guszejnov:2018.isothermal}
{Guszejnov} D.,  {Hopkins} P.~F.,  {Grudi{\'c}} M.~Y.,  {Krumholz} M.~R.,
  {Federrath} C.,  2018, \mn@doi [\mnras] {10.1093/mnras/sty1847}, \href
  {http://adsabs.harvard.edu/abs/2018MNRAS.480..182G} {480, 182}

\bibitem[\protect\citeauthoryear{{Hills}}{{Hills}}{1980}]{hills:1980}
{Hills} J.~G.,  1980, \mn@doi [\apj] {10.1086/157703}, \href
  {http://adsabs.harvard.edu/abs/1980ApJ...235..986H} {235, 986}

\bibitem[\protect\citeauthoryear{{Hopkins}}{{Hopkins}}{2015}]{hopkins:gizmo}
{Hopkins} P.~F.,  2015, \mn@doi [\mnras] {10.1093/mnras/stv195}, \href
  {http://adsabs.harvard.edu/abs/2015MNRAS.450...53H} {450, 53}

\bibitem[\protect\citeauthoryear{{Hopkins} \& {Grudi\'{c}}}{{Hopkins} \&
  {Grudi\'{c}}}{2018}]{hopkins.grudic:2018.rp}
{Hopkins} P.~F.,  {Grudi\'{c}} M.~Y.,  2018, preprint, \href
  {http://adsabs.harvard.edu/abs/2018arXiv180307573H} {} (\mn@eprint {arXiv}
  {1803.07573})

\bibitem[\protect\citeauthoryear{{Hopkins}, {Quataert}  \& {Murray}}{{Hopkins}
  et~al.}{2011}]{hopkins:2011.mom.feedback}
{Hopkins} P.~F.,  {Quataert} E.,   {Murray} N.,  2011, \mn@doi [\mnras]
  {10.1111/j.1365-2966.2011.19306.x}, \href
  {http://adsabs.harvard.edu/abs/2011MNRAS.417..950H} {417, 950}

\bibitem[\protect\citeauthoryear{{Hopkins}, {Quataert}  \& {Murray}}{{Hopkins}
  et~al.}{2012}]{hopkins:fb.ism.prop}
{Hopkins} P.~F.,  {Quataert} E.,   {Murray} N.,  2012, \mn@doi [\mnras]
  {10.1111/j.1365-2966.2012.20578.x}, \href
  {http://adsabs.harvard.edu/abs/2012MNRAS.421.3488H} {421, 3488}

\bibitem[\protect\citeauthoryear{{Hopkins} et~al.,}{{Hopkins}
  et~al.}{2018}]{fire2}
{Hopkins} P.~F.,  et~al., 2018, \mn@doi [\mnras] {10.1093/mnras/sty1690}, \href
  {http://adsabs.harvard.edu/abs/2018MNRAS.480..800H} {480, 800}

\bibitem[\protect\citeauthoryear{{Howard}, {Pudritz}  \& {Harris}}{{Howard}
  et~al.}{2016}]{howard:2016}
{Howard} C.~S.,  {Pudritz} R.~E.,   {Harris} W.~E.,  2016, \mn@doi [\mnras]
  {10.1093/mnras/stw1476}, \href
  {http://adsabs.harvard.edu/abs/2016MNRAS.461.2953H} {461, 2953}

\bibitem[\protect\citeauthoryear{{Howard}, {Pudritz}  \& {Harris}}{{Howard}
  et~al.}{2017}]{howard:2017}
{Howard} C.~S.,  {Pudritz} R.~E.,   {Harris} W.~E.,  2017, \mn@doi [\mnras]
  {10.1093/mnras/stx1363}, \href
  {http://adsabs.harvard.edu/abs/2017MNRAS.470.3346H} {470, 3346}

\bibitem[\protect\citeauthoryear{{Hu}, {Naab}, {Walch}, {Glover}  \&
  {Clark}}{{Hu} et~al.}{2016}]{hu:2016}
{Hu} C.-Y.,  {Naab} T.,  {Walch} S.,  {Glover} S.~C.~O.,   {Clark} P.~C.,
  2016, \mn@doi [\mnras] {10.1093/mnras/stw544}, \href
  {http://adsabs.harvard.edu/abs/2016MNRAS.458.3528H} {458, 3528}

\bibitem[\protect\citeauthoryear{Hunter}{Hunter}{2007}]{matplotlib}
Hunter J.~D.,  2007, \mn@doi [Computing In Science \& Engineering]
  {10.1109/MCSE.2007.55}, 9, 90

\bibitem[\protect\citeauthoryear{Jones, Oliphant, Peterson  et~al.}{Jones
  et~al.}{2001}]{scipy}
Jones E.,  Oliphant T.,  Peterson P.,   et~al., 2001, {SciPy}: Open source
  scientific tools for {Python}, \url {http://www.scipy.org/}

\bibitem[\protect\citeauthoryear{{Jumper} \& {Matzner}}{{Jumper} \&
  {Matzner}}{2018}]{jumper.matzner:2018.rp}
{Jumper} P.~H.,  {Matzner} C.~D.,  2018, \mn@doi [\mnras]
  {10.1093/mnras/sty1784}, \href
  {http://adsabs.harvard.edu/abs/2018MNRAS.480..905J} {480, 905}

\bibitem[\protect\citeauthoryear{{Keller}, {Wadsley}, {Benincasa}  \&
  {Couchman}}{{Keller} et~al.}{2014}]{keller:2014.superbubbles}
{Keller} B.~W.,  {Wadsley} J.,  {Benincasa} S.~M.,   {Couchman} H.~M.~P.,
  2014, \mn@doi [\mnras] {10.1093/mnras/stu1058}, \href
  {http://adsabs.harvard.edu/abs/2014MNRAS.442.3013K} {442, 3013}

\bibitem[\protect\citeauthoryear{{Kim}, {Kim}  \& {Ostriker}}{{Kim}
  et~al.}{2016}]{kim:2016.dusty.hii}
{Kim} J.-G.,  {Kim} W.-T.,   {Ostriker} E.~C.,  2016, \mn@doi [\apj]
  {10.3847/0004-637X/819/2/137}, \href
  {https://ui.adsabs.harvard.edu/#abs/2016ApJ...819..137K} {819, 137}

\bibitem[\protect\citeauthoryear{{Kim}, {Kim}, {Ostriker}  \& {Skinner}}{{Kim}
  et~al.}{2017}]{kim:2017.rhd}
{Kim} J.-G.,  {Kim} W.-T.,  {Ostriker} E.~C.,   {Skinner} M.~A.,  2017, \mn@doi
  [\apj] {10.3847/1538-4357/aa9b80}, \href
  {http://adsabs.harvard.edu/abs/2017ApJ...851...93K} {851, 93}

\bibitem[\protect\citeauthoryear{{Kim} et~al.,}{{Kim}
  et~al.}{2018a}]{kim:2018.fire.gcs}
{Kim} J.-h.,  et~al., 2018a, \mn@doi [\mnras] {10.1093/mnras/stx2994}, \href
  {http://adsabs.harvard.edu/abs/2018MNRAS.474.4232K} {474, 4232}

\bibitem[\protect\citeauthoryear{{Kim}, {Kim}  \& {Ostriker}}{{Kim}
  et~al.}{2018b}]{kim:2018}
{Kim} J.-G.,  {Kim} W.-T.,   {Ostriker} E.~C.,  2018b, \mn@doi [\apj]
  {10.3847/1538-4357/aabe27}, \href
  {http://adsabs.harvard.edu/abs/2018ApJ...859...68K} {859, 68}

\bibitem[\protect\citeauthoryear{{Klessen}, {Heitsch}  \& {Mac Low}}{{Klessen}
  et~al.}{2000}]{klessen:2000.cloud.collapse}
{Klessen} R.~S.,  {Heitsch} F.,   {Mac Low} M.-M.,  2000, \mn@doi [\apj]
  {10.1086/308891}, \href {http://adsabs.harvard.edu/abs/2000ApJ...535..887K}
  {535, 887}

\bibitem[\protect\citeauthoryear{{Kroupa}}{{Kroupa}}{2002}]{kroupa:imf}
{Kroupa} P.,  2002, \mn@doi [Science] {10.1126/science.1067524}, \href
  {http://adsabs.harvard.edu/abs/2002Sci...295...82K} {295, 82}

\bibitem[\protect\citeauthoryear{{Kruijssen}}{{Kruijssen}}{2012}]{kruijssen:2012.cluster.formation.efficiency}
{Kruijssen} J.~M.~D.,  2012, \mn@doi [\mnras]
  {10.1111/j.1365-2966.2012.21923.x}, \href
  {http://adsabs.harvard.edu/abs/2012MNRAS.426.3008K} {426, 3008}

\bibitem[\protect\citeauthoryear{{Krumholz}}{{Krumholz}}{2018}]{krumholz:2018.rp}
{Krumholz} M.~R.,  2018, \mn@doi [\mnras] {10.1093/mnras/sty2105}, \href
  {http://adsabs.harvard.edu/abs/2018MNRAS.480.3468K} {480, 3468}

\bibitem[\protect\citeauthoryear{{Krumholz}, {Klein}  \& {McKee}}{{Krumholz}
  et~al.}{2011}]{krumholz:2011.rhd.starcluster.sim}
{Krumholz} M.~R.,  {Klein} R.~I.,   {McKee} C.~F.,  2011, \mn@doi [\apj]
  {10.1088/0004-637X/740/2/74}, \href
  {http://adsabs.harvard.edu/abs/2011arXiv1104.2038K} {740, 74}

\bibitem[\protect\citeauthoryear{{Krumholz} et~al.,}{{Krumholz}
  et~al.}{2014}]{krumholz:2014.feedback.review}
{Krumholz} M.~R.,  et~al., 2014, \mn@doi [Protostars and Planets VI]
  {10.2458/azu_uapress_9780816531240-ch011}, \href
  {http://adsabs.harvard.edu/abs/2014prpl.conf..243K} {pp 243--266}

\bibitem[\protect\citeauthoryear{Lam, Pitrou  \& Seibert}{Lam
  et~al.}{2015}]{numba}
Lam S.~K.,  Pitrou A.,   Seibert S.,  2015, in Proceedings of the Second
  Workshop on the LLVM Compiler Infrastructure in HPC. LLVM '15.
ACM, New York, NY, USA, pp 7:1--7:6, \mn@doi{10.1145/2833157.2833162}, \url
  {http://doi.acm.org/10.1145/2833157.2833162}

\bibitem[\protect\citeauthoryear{{Leitherer} et~al.}{{Leitherer}
  et~al.}{1999}]{starburst99}
{Leitherer} C.,  et~al., 1999, \mn@doi [\apjs] {10.1086/313233}, \href
  {http://adsabs.harvard.edu/abs/1999ApJS..123....3L} {123, 3}

\bibitem[\protect\citeauthoryear{{McKee} \& {Ostriker}}{{McKee} \&
  {Ostriker}}{2007}]{mckee:2007.review}
{McKee} C.~F.,  {Ostriker} E.~C.,  2007, \mn@doi [\araa]
  {10.1146/annurev.astro.45.051806.110602}, \href
  {http://adsabs.harvard.edu/abs/2007ARA%26A..45..565M} {45, 565}

\bibitem[\protect\citeauthoryear{{Murray}, {Quataert}  \& {Thompson}}{{Murray}
  et~al.}{2010}]{murray:molcloud.disrupt.by.rad.pressure}
{Murray} N.,  {Quataert} E.,   {Thompson} T.~A.,  2010, \mn@doi [\apj]
  {10.1088/0004-637X/709/1/191}, \href
  {http://adsabs.harvard.edu/abs/2009arXiv0906.5358M} {709, 191}

\bibitem[\protect\citeauthoryear{{Naab} \& {Ostriker}}{{Naab} \&
  {Ostriker}}{2017}]{naab.ostriker:2017.review}
{Naab} T.,  {Ostriker} J.~P.,  2017, \mn@doi [\araa]
  {10.1146/annurev-astro-081913-040019}, \href
  {http://adsabs.harvard.edu/abs/2017ARA%26A..55...59N} {55, 59}

\bibitem[\protect\citeauthoryear{{Offner}, {Clark}, {Hennebelle}, {Bastian},
  {Bate}, {Hopkins}, {Moraux}  \& {Whitworth}}{{Offner}
  et~al.}{2014}]{offner:2014.imf.review}
{Offner} S.~S.~R.,  {Clark} P.~C.,  {Hennebelle} P.,  {Bastian} N.,  {Bate}
  M.~R.,  {Hopkins} P.~F.,  {Moraux} E.,   {Whitworth} A.~P.,  2014, \mn@doi
  [Protostars and Planets VI] {10.2458/azu_uapress_9780816531240-ch003}, \href
  {http://adsabs.harvard.edu/abs/2014prpl.conf...53O} {pp 53--75}

\bibitem[\protect\citeauthoryear{{Orr} et~al.,}{{Orr}
  et~al.}{2018}]{orr:2018.kennicutt.schmidt}
{Orr} M.~E.,  et~al., 2018, \mn@doi [\mnras] {10.1093/mnras/sty1241}, \href
  {http://adsabs.harvard.edu/abs/2018MNRAS.478.3653O} {478, 3653}

\bibitem[\protect\citeauthoryear{P\'erez \& Granger}{P\'erez \&
  Granger}{2007}]{ipython}
P\'erez F.,  Granger B.~E.,  2007, \mn@doi [Computing in Science and
  Engineering] {10.1109/MCSE.2007.53}, 9, 21

\bibitem[\protect\citeauthoryear{{Price} \& {Bate}}{{Price} \&
  {Bate}}{2008}]{price.bate:2008.magnetic.fields}
{Price} D.~J.,  {Bate} M.~R.,  2008, \mn@doi [\mnras]
  {10.1111/j.1365-2966.2008.12976.x}, \href
  {http://adsabs.harvard.edu/abs/2008MNRAS.385.1820P} {385, 1820}

\bibitem[\protect\citeauthoryear{{Raskutti}, {Ostriker}  \&
  {Skinner}}{{Raskutti} et~al.}{2016}]{raskutti:2016.gmcs}
{Raskutti} S.,  {Ostriker} E.~C.,   {Skinner} M.~A.,  2016, \mn@doi [\apj]
  {10.3847/0004-637X/829/2/130}, \href
  {http://adsabs.harvard.edu/abs/2016ApJ...829..130R} {829, 130}

\bibitem[\protect\citeauthoryear{{Rosdahl} \& {Teyssier}}{{Rosdahl} \&
  {Teyssier}}{2015}]{rosdahl:2015.rhd}
{Rosdahl} J.,  {Teyssier} R.,  2015, \mn@doi [\mnras] {10.1093/mnras/stv567},
  \href {http://adsabs.harvard.edu/abs/2015MNRAS.449.4380R} {449, 4380}

\bibitem[\protect\citeauthoryear{{Skinner} \& {Ostriker}}{{Skinner} \&
  {Ostriker}}{2013}]{skinner:2013.hyperion}
{Skinner} M.~A.,  {Ostriker} E.~C.,  2013, \mn@doi [\apjs]
  {10.1088/0067-0049/206/2/21}, \href
  {http://adsabs.harvard.edu/abs/2013ApJS..206...21S} {206, 21}

\bibitem[\protect\citeauthoryear{{Skinner} \& {Ostriker}}{{Skinner} \&
  {Ostriker}}{2015}]{skinner:2015.ir.molcloud.disrupt}
{Skinner} M.~A.,  {Ostriker} E.~C.,  2015, \mn@doi [\apj]
  {10.1088/0004-637X/809/2/187}, \href
  {http://adsabs.harvard.edu/abs/2015ApJ...809..187S} {809, 187}

\bibitem[\protect\citeauthoryear{{Sormani}, {Tre{\ss}}, {Klessen}  \&
  {Glover}}{{Sormani} et~al.}{2017}]{sormani:2016.imf.sampling}
{Sormani} M.~C.,  {Tre{\ss}} R.~G.,  {Klessen} R.~S.,   {Glover} S.~C.~O.,
  2017, \mn@doi [\mnras] {10.1093/mnras/stw3205}, \href
  {http://adsabs.harvard.edu/abs/2017MNRAS.466..407S} {466, 407}

\bibitem[\protect\citeauthoryear{{Su} et~al.,}{{Su}
  et~al.}{2018}]{su:2017.discreteness}
{Su} K.-Y.,  et~al., 2018, \mn@doi [\mnras] {10.1093/mnras/sty1928}, \href
  {http://adsabs.harvard.edu/abs/2018MNRAS.480.1666S} {480, 1666}

\bibitem[\protect\citeauthoryear{V\'{a}zquez-Semadeni, Col\'{i}n, G\'{o}mez,
  Ballesteros-Paredes  \& Watson}{V\'{a}zquez-Semadeni
  et~al.}{2010}]{vazquez:2010}
V\'{a}zquez-Semadeni E.,  Col\'{i}n P.,  G\'{o}mez G.~C.,  Ballesteros-Paredes
  J.,   Watson A.~W.,  2010, The Astrophysical Journal, 715, 1302

\bibitem[\protect\citeauthoryear{{V{\'a}zquez-Semadeni},
  {Gonz{\'a}lez-Samaniego}  \& {Col{\'{\i}}n}}{{V{\'a}zquez-Semadeni}
  et~al.}{2017}]{vazquez:2017}
{V{\'a}zquez-Semadeni} E.,  {Gonz{\'a}lez-Samaniego} A.,   {Col{\'{\i}}n} P.,
  2017, \mn@doi [\mnras] {10.1093/mnras/stw3229}, \href
  {http://adsabs.harvard.edu/abs/2017MNRAS.467.1313V} {467, 1313}

\bibitem[\protect\citeauthoryear{{da Silva}, {Fumagalli}  \& {Krumholz}}{{da
  Silva} et~al.}{2012}]{slug}
{da Silva} R.~L.,  {Fumagalli} M.,   {Krumholz} M.,  2012, \mn@doi [\apj]
  {10.1088/0004-637X/745/2/145}, \href
  {https://ui.adsabs.harvard.edu/#abs/2012ApJ...745..145D} {745, 145}

\makeatother
\end{thebibliography}







\bsp	
\label{lastpage}
\end{document}